\documentclass[aps,prl,preprint,preprintnumbers,amsmath,amssymb]{revtex4}

\usepackage{graphicx}
\usepackage[colorlinks=true,citecolor=blue]{hyperref}

\begin{document}
    \title{Coherent Optical Spectroscopy of a Single Quantum Dot Via an Off-Resonant Cavity}
    \author{Arka Majumdar$^1$}
    \email{arkam@stanford.edu}
    \author{Alexander Papageorge$^1$}
    \author{Erik D. Kim$^1$}
    \author{Michal Bajscy$^1$}
    \author{Hyochul Kim$^2$}
    \author{Pierre Petroff$^2$}
    \author{Jelena Vu\v{c}kovi\'{c}$^1$}
    \affiliation{$^1$E.L.Ginzton Laboratory, Stanford University, Stanford, CA, 94305\\}
    \affiliation{$^2$Materials Department, University of California, Santa Barbara, CA
    93106\\}
\begin{abstract}
In recent experiments on coupled quantum dot (QD) optical cavity
systems a pronounced interaction between the dot and the cavity
has been observed even for detunings of many cavity linewidths.
This interaction has been attributed to an incoherent
phonon-mediated scattering process and is absent in atomic
systems. Here, we demonstrate that despite its incoherent nature,
this process preserves the signatures of coherent interaction
between a QD and a strong driving laser, which may be observed via
the optical emission from the off-resonant cavity. Under
bichromatic driving of the QD, the cavity emission exhibits
spectral features consistent with optical dressing of the QD
transition. In addition to revealing new aspects of the
off-resonant QD-cavity interaction, this result provides a new,
simpler means of coherently probing QDs than traditional
approaches and opens the possibility of employing off-resonant
cavities to optically interface QD-nodes in quantum networks.
\end{abstract}
\maketitle

Optically controlled quantum dot (QD) spins coupled to
semiconductor microcavities constitute a promising platform for
robust and scalable quantum information processing devices, where
QD spin nodes are optically interconnected via photonic circuits.
As such, much effort in recent years has been dedicated to
demonstrating fast optical control of a QD spin
\cite{article:press08,article_Erik_Spin} and to studying QD-
cavity quantum electrodynamics (CQED) phenomena
\cite{article:eng07}. The prospect of strongly enhanced
light-matter interactions between a QD and an optical field has
served as a focal impetus in integrating QDs with high quality
factor (Q) optical cavities, with maximum enhancement occurring
when the QD and the cavity are resonant and the QD is spatially
aligned to the cavity mode. Since achieving this maximum
enhancement is difficult due to limitations in growth and
fabrication techniques, the recently observed coupling between a
single QD and a detuned optical cavity mode
\cite{article:majumdar09,article:michler09} has spurred
considerable theoretical \cite{article:electron_phonon_cqed,
majumdar_phonon_11} and experimental interest in determining the
physical mechanism behind such coupling as well as in possible
applications. Though recent experiments have investigated the
linewidth and saturation behavior of this off-resonant cavity
emission \cite{article:majumdar10, article:michler10}, relatively
little has been done to investigate the potential utility of such
measurements in performing coherent optical spectroscopy of single
QDs.

\begin{figure}
\centering
\includegraphics{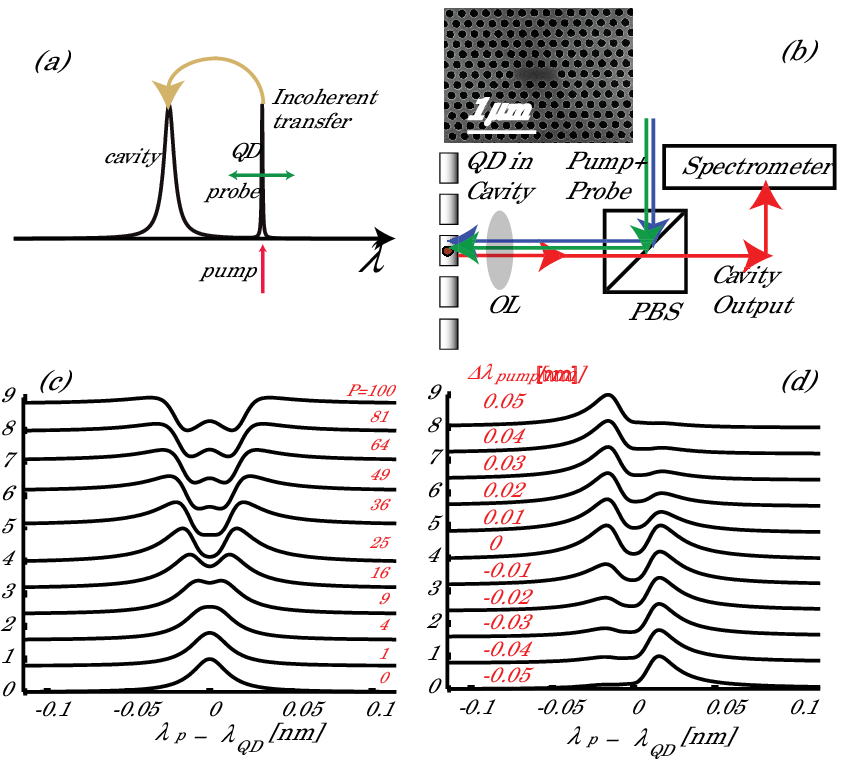}
\caption{ Experimental setup and numerical simulations. (a) The
schematic shows the relative position of the QD and the cavity on
a wavelength axis. For the particular QD-cavity system considered,
the QD is red detuned from the cavity, though off-resonant
coupling is observed for both red and blue detuned QDs. In
experiments, a strong pump laser dresses the QD while a weak probe
laser is scanned across the QD. QD emission is incoherently
coupled to the cavity. The cavity emission is monitored as a
function of probe laser wavelength $\lambda_p$. (b) The
experimental setup is a confocal cross-polarization setup. The PBS
(polarizing beam splitter) is used to perform cross-polarized
reflectivity measurements, as in previous work
\cite{article:eng07}. The powers are measured in front of the
objective lens (OL). The output is dispersed in a single-grating
monochromator and measured by a nitrogen-cooled CCD. We employ a
linear three hole defect PC cavity (a scanning electron micrograph
is shown in the inset). (c) Normalized off-resonant cavity
emission obtained by numerical simulation is plotted as a function
of $\lambda_p-\lambda_{QD}$, $\lambda_{QD}$ and $\lambda_p$ being
the QD resonance and probe laser wavelengths, respectively, for
different pump powers $P$ (normalized units) while the probe power
is kept at $1$. The pump laser is resonant with the QD. (d) For a
pump power of $P=25$, the cavity emission is plotted as a function
of $\lambda_p-\lambda_{QD}$ for different pump-QD detunings
$\Delta\lambda_{pump}$ (nm). In both (c) and (d), spectra are
vertically offset for clarity. \label{Fig1_theory}}
\end{figure}

Here, we present both theoretical and experimental studies of a
strongly-driven QD that is off-resonantly coupled to a photonic
crystal (PC) cavity mode. In these studies, a strong
narrow-bandwidth CW pump laser serves to dress the QD, while a
weaker continuous wave (CW) probe laser is scanned across the QD
resonance; the output signal is always collected at the spectrally
detuned cavity (Fig. \ref{Fig1_theory} a). We model the
bichromatic driving of the QD coupled to an off-resonant cavity by
adding an incoherent phonon-mediated coupling between the QD and
the cavity and perform simulations with realistic system
parameters. The bichromatic driving of a two-level system has been
analyzed before \cite{bichromatic_QD}. We use similar techniques
to analyze the driving of a two-level system such as a QD,
incoherently coupled to an off-resonant cavity via phonons
\cite{majumdar_phonon_11} (see Supplementary Material). In these
simulations we neglect any coherent coupling between the QD and
the cavity (i.e., the vacuum Rabi splitting $g = 0$). Figure
\ref{Fig1_theory} c shows the theoretically calculated cavity
output as a function of the probe laser wavelength $\lambda_p$ for
different powers $P$ of the resonant pump laser. At low pump
power, we observe a simple Lorentzian line-shape with QD linewidth
\cite{article:majumdar10}. However, as the pump power is
increased, the Lorentzian peak splits into two peaks, the
separation between the peaks increasing linearly with pump Rabi
frequency. We find that these two peaks are separated by $\sim 4$
times the Rabi frequency (see Supplementary material). As the pump
power is increased further, a third peak corresponding to the
central Mollow peak appears at the QD resonance, leading to the
emergence of two dips whose separation also increases linearly
with pump Rabi frequency. We note that the lack of a prominent
central Mollow peak as observed in resonance fluorescence studies
of single QDs \cite{article:shih09, article:nick09} is a result of
the saturation of the QD absorption and, hence, of the cavity
emission. As such, these cavity emission measurements are more
akin to absorption measurements of a strongly driven QD
\cite{Xu:QD_spectrscopy_2007_science} rather than the
aforementioned resonance fluorescence measurements
\cite{article:shih09,article:nick09}. Figure \ref{Fig1_theory} d
plots the cavity output for different detunings
$\Delta\lambda_{pump}=\lambda_{pump}-\lambda_{QD}$ between the
pump and the QD. We observe that the two peaks remain distinct but
become asymmetric when the pump is detuned from the QD. This is
consistent with the anti-crossing of the Rabi sidebands of the
dressed QD that occurs as the pump is tuned through the QD
resonance \cite{Xu:PRL_Mollow}. These theoretical results
demonstrate that measurements of cavity emission allow for the
observation of phenomena associated with the coherent optical
driving of the QD.

\begin{figure}
\centering
\includegraphics{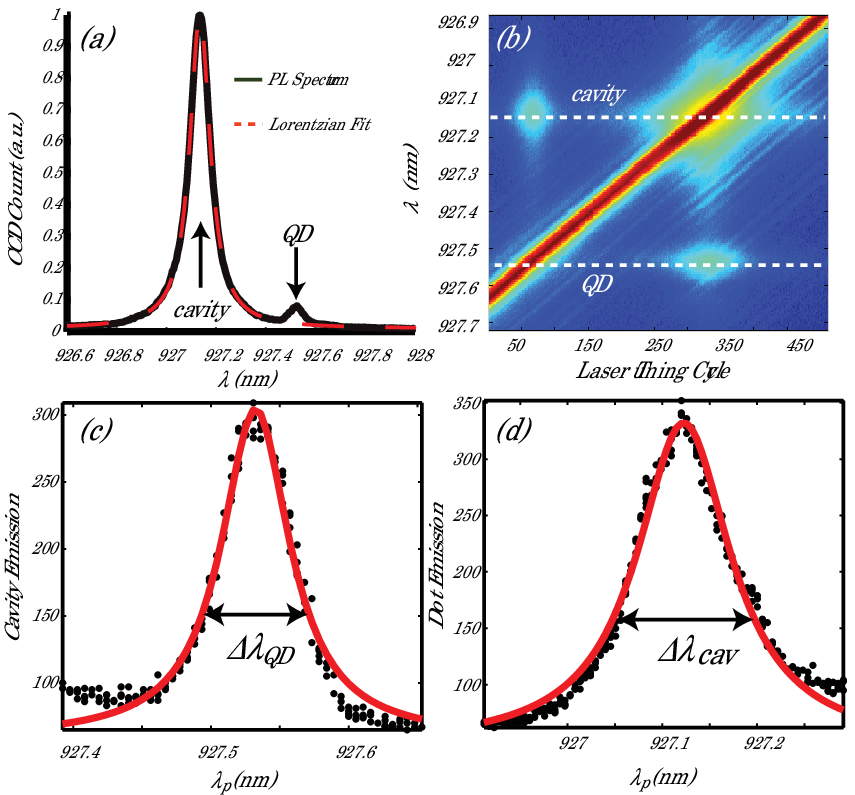}
\caption{Characterization of the QD-cavity system in
photoluminescence (PL) and probing of the off-resonant dot-cavity
coupling. (a) PL spectrum of the system. From the Lorentzian fit
to the cavity we estimate a cavity linewidth
$\Delta\lambda_{cav}=0.1$ nm. (b) The laser is scanned across the
QD-cavity system. Emission from the cavity is observed when the
laser is resonant with the QD. Similarly, emission from the QD is
observed when the laser is resonant with the cavity. (c),(d) The
QD (cavity) linewidth is measured by monitoring the cavity (QD)
emission as a function of the probe wavelength $\lambda_p$.
\label{Fig2_system_character} }
\end{figure}

To demonstrate the use of such cavity emission to perform coherent
optical spectroscopy of an off-resonantly coupled QD, we perform a
series of experiments measuring the optical emission spectra of a
system consisting of a single self-assembled InAs QD
off-resonantly coupled to a linear three hole defect GaAs PC
cavity under different optical excitation configurations in a
helium-flow cryostat at cryogenic temperatures ($\sim 30-35$ K)
(Fig. \ref{Fig1_theory} b) \cite{article:eng07}. The $160$nm GaAs
membrane used to fabricate the photonic crystal is grown by
molecular beam epitaxy on top of a GaAs $(100)$ wafer. A low
density layer of InAs QDs is grown in the center of the membrane
($80$ nm beneath the surface). The GaAs membrane sits on a $918$
nm sacrificial layer of Al$_{0.8}$Ga$_{0.2}$As. Under the
sacrificial layer, a $10$-period distributed Bragg reflector,
consisting of a quarter-wave AlAs/GaAs stack, is used to increase
collection into the objective lens. The photonic crystal was
fabricated using electron beam lithography, dry plasma etching,
and wet etching of the sacrificial layer in diluted hydrofluoric
acid, as described previously \cite{article:eng07}. Optical
emission is collected and dispersed by a single grating
monochromator and then measured by a liquid nitrogen cooled charge
coupled device (CCD). We first characterize the coupled QD-cavity
system by measuring the photoluminescence (PL) spectrum obtained
under above-band excitation by an $820$ nm Ti:sapphire laser (Fig.
\ref{Fig2_system_character} a). From the Lorentzian fit to the
cavity resonance, we find that the cavity linewidth is
$\Delta\lambda_{cav}=0.1$ nm, corresponding to a cavity field
decay rate of $\kappa/2\pi=17$ GHz. We do not observe the
anti-crossing of the cavity and QD peaks when the QD is tuned
across the cavity resonance by changing temperature, indicating
that the QD is not strongly coupled to the cavity. The QD
resonance is at $\lambda_{QD}=927.5$ nm and the cavity resonance
is at $\lambda_{cav}=927.1$ nm at $35$ K temperature leading to a
dot-cavity detuning $\Delta\lambda=\lambda_{QD}-\lambda_{cav}=0.4$
nm. When we scan a laser across the off-resonant QD-cavity system
we observe cavity emission when the laser is resonant with the QD
and QD emission when the laser is resonant with the cavity (Fig.
\ref{Fig2_system_character} b). We can estimate the linewidth of
the QD ($\Delta\lambda_{QD}= 0.06$ nm) and the cavity
($\Delta\lambda_{cav}=0.11$ nm) by scanning the excitation laser
across one resonance and observing emission at the other (Fig.
\ref{Fig2_system_character} c and d). These measurements yield a
broader cavity linewidth compared to that measured in standard PL
measurements due to the heating of the structure caused by the
resonant laser \cite{article:majumdar10}. We now use this
off-resonant cavity emission to probe the dressing of the QD by a
strong resonant laser field.

\begin{figure}
\centering
\includegraphics[width=3in]{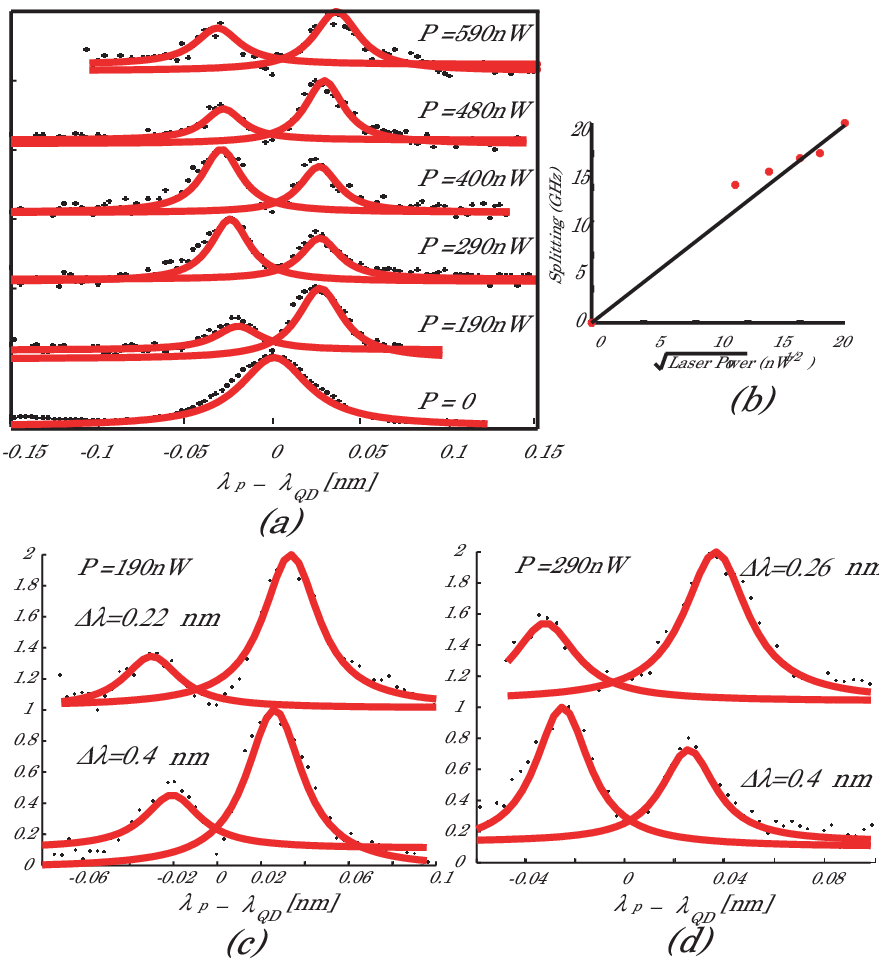}
\caption{Coherent interaction between the QD and the laser
observed through cavity emission. (a) Normalized cavity emission
as a function of the probe laser wavelength for different pump
powers (measured before the objective lens). We observe that a
single QD resonance splits into two peaks. The splitting is
linearly proportional to the Rabi frequency of the pump laser.
Each peak is fit with a Lorentzian. (b) Rabi frequency $\Omega$ of
the laser (estimated from the splitting) as a function of the
square root of the pump power $P$. A linear relation exists
between $\Omega$ and $\sqrt{P}$. (c) Normalized cavity emission
for a pump power of $190$ nW for two different QD-cavity detunings
$\Delta\lambda=0.22$ and $0.4$ nm. (d) Cavity emission for a pump
power of $290$ nW at two different QD-cavity detunings
$\Delta\lambda=0.26$ and $0.4$ nm. We observe that the splitting
increases for smaller detuning (i.e., when the pump laser is
closer to the cavity), which suggests that the input laser power
is enhanced by the presence of the cavity. For all experiments the
probe laser power is kept constant at $20$ nW. The QD-cavity
detuning is defined as $\Delta\lambda=\lambda_{QD}-\lambda_{cav}$.
In (a), (c), (d) the spectra are vertically offset for clarity.
\label{Figure3_power_dependence}}
\end{figure}

First, we experimentally investigate bi-chromatic driving of the
off-resonantly coupled QD-cavity system under the same conditions
as modeled in Fig. \ref{Fig1_theory} a: a strong pump laser is
used to resonantly drive the QD, while a weak probe laser is used
to scan across the QD resonance. For these experiments, we utilize
the spectral selectivity provided by the experimental setup to
isolate the cavity emission and measure it as a function of the
probe wavelength. A cross-polarized confocal microscopy setup is
employed in this case (Fig. \ref{Fig1_theory} b), although
cross-polarization is not essential here as the QD and the cavity
emission are co-polarized. Fig. \ref{Figure3_power_dependence} a
shows the cavity emission intensity as a function of the probe
laser wavelength $\lambda_p$. In the absence of the pump laser
$(P=0)$, we observe that the cavity emission spectrum possesses a
Lorentzian line-shape. However, when a strong pump drives the QD,
the Lorentzian splits into two peaks, as observed in the
simulations in Fig. \ref{Fig1_theory} c. However experimentally
measured QD linewidths are broadened by spectral diffusion of the
QD transition, which is not included in our theoretical model
\cite{fluctuating_environment_PRB}. Hence, we fit a Lorentzian to
each peak and study the splitting between two peaks as a function
of the pump laser power. Fig. \ref{Figure3_power_dependence} b
plots this splitting as a function of the square root of the laser
power $P$ measured in front of the objective lens (OL). We observe
that the splitting increases linearly with $\sqrt{P}\propto E$,
the laser field amplitude. The splitting is given by $\sim 4$
times the laser Rabi frequency $\Omega=\mu_dE/\hbar$, where
$\mu_d$ is the QD dipole moment. We note that in the results of
Fig. \ref{Figure3_power_dependence} a, the peaks are are not
symmetric, due to the fact that fixing the pump laser exactly to
the QD resonance in experiments is made difficult by spectral
drifts in both the QD resonance and the pump laser wavelength over
time. For a detuned pump, the splitting is modified, and this
causes a deviation of the Rabi frequencies from the linear
relation as shown in Fig. \ref{Figure3_power_dependence} b. We
also note that the high pump power regime of Fig.
\ref{Fig1_theory} c, which shows a central peak and two dips in
the observed spectra, is difficult to observe in experiments due
to the fact that the CCD also collects transmitted pump light. At
higher powers, this transmitted pump light can saturate CCD pixels
corresponding to wavelengths near the pump wavelength. This
saturation can result in charge leakage across CCD pixels leading
to a deterioration of the signal to noise ratio of cavity emission
measurements. The use of improved spectral filtering techniques
would reduce the amount of pump light collected, possibly enabling
observation of this high power regime.

We estimate that the off-resonant cavity $(\Delta\lambda=0.4 nm)$
enhances the laser electric field inside cavity by a factor of
$\sim 40$, compared to the bare QD case, assuming a spot size of
$3 \mu$m and QD at the field maximum (see supplement). This agrees
with the result shown in Fig. \ref{Figure3_power_dependence} c,
where the cavity emission is plotted for two different QD-cavity
detunings at the same pump power. The Rabi frequencies of the
laser at a QD-cavity detuning of $\Delta\lambda=0.4$ nm are
measured to be $8.15$ and $8.9$ GHz at input powers of $190$ and
$290$ nW, respectively. The Rabi frequencies increase to $11.1$
and $12.1$ GHz when the pump is closer to cavity ($\Delta\lambda$
are $0.26$ and $0.22$ nm, respectively). We theoretically estimate
these Rabi frequencies to be $11.6$ and $13.8$ GHz, which are
close to the experimentally measured values.

\begin{figure}
\centering
\includegraphics[width=2.5in]{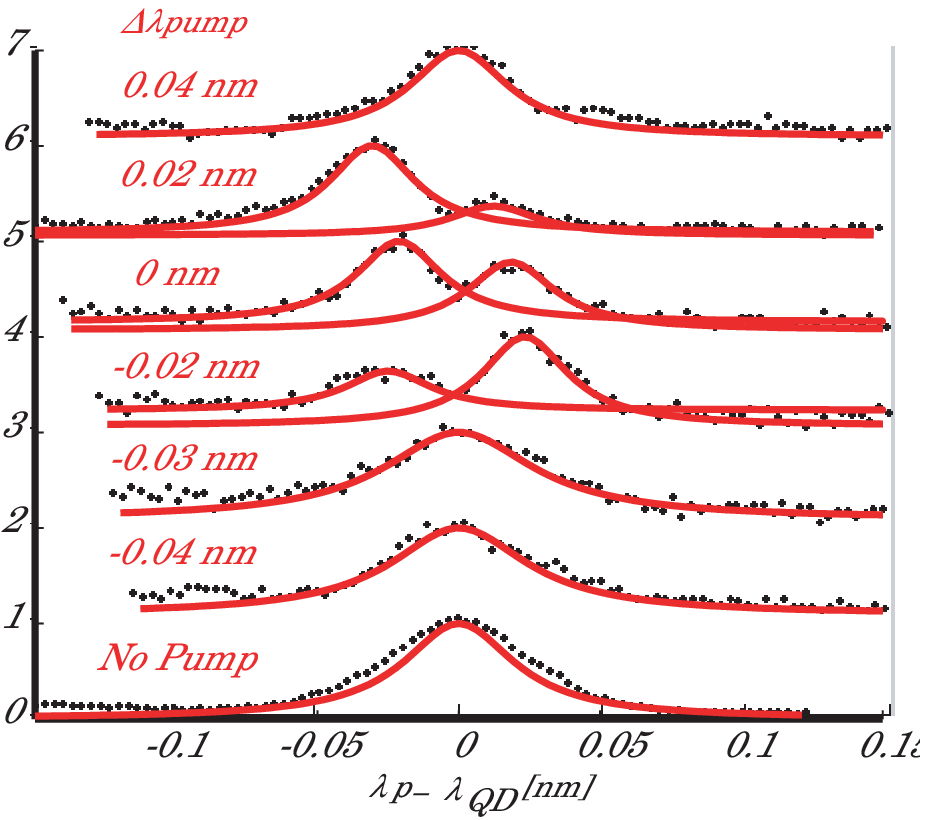}
\caption{ Dependence of the result on pump-QD detuning.
Off-resonant cavity emission as a function of the probe laser
wavelength for different pump-QD detunings
$\Delta\lambda_{pump}=\lambda_{pump}-\lambda_{QD}$. We observe
that the QD linewidth broadens when the pump is present and
detuned from the QD resonance. As the pump is tuned through the QD
resonance, we observe the emergence of two peaks in the cavity
emission spectrum. This two-peak spectrum is consistent with the
observation of the anti-crossing of Rabi sidebands. The pump and
probe power are kept at $290$ nW and $20$ nW, respectively. The
spectra are offset for clarity.
\label{Figure4_detuning_dependence}}
\end{figure}

Finally, we study the effects of the detuning between the pump and
the QD resonance on the off-resonant cavity emission. Fig.
\ref{Figure4_detuning_dependence} shows the cavity emission as a
function of probe laser wavelength $\lambda_p$ for different pump
laser-QD detunings
$\Delta\lambda_{pump}=\lambda_{pump}-\lambda_{QD}$. The pump laser
power is kept fixed at $290$ nW. The detuning
$\Delta\lambda_{pump}$ is changed from $-0.04$ nm (blue detuned)
to $0.04$ nm (red detuned). We observe that when the pump laser is
far detuned from the QD resonance, the cavity emission shows a
single peak with $\lambda_p$. As the pump is tuned closer to the
cavity resonance, two peaks emerge in the spectrum, where the
peaks are asymmetric when the pump is not exactly resonant with
the QD. The fact that the peaks remain distinct as the pump is
tuned through the QD resonance verifies experimental observation
of the anti-crossing of the Rabi sidebands of the driven QD,
consistent with the theory (Fig. \ref{Fig1_theory} d).

In conclusion, we demonstrate that signatures of the coherent
driving of a QD by a strong pump laser are preserved after phonon
assisted scattering to an off-resonant cavity despite the fact
that this scattering process is incoherent. In addition to
revealing new aspects of the off-resonant QD-cavity interaction,
this result is also potentially useful for enabling simpler
coherent optical spectroscopy of a QD, as the readout signal is
offset in frequency and can be spectrally filtered using
well-established techniques. Moreover, this approach may relax the
requirement of working exclusively with strongly coupled QD-cavity
systems in quantum networks.

\section{Supplementary}
\section{Estimation of electric field enhancement:}
We consider a Gaussian laser beam with power $P$ and frequency
$\omega$ incident on a photonic crystal cavity. The power is
measured in front of the objective lens and the coupling
efficiency of the laser to the cavity is $\eta$. If the cavity
quality factor is $Q=\omega_0/\Delta\omega$, with cavity resonance
frequency $\omega_0$ and linewidth $\Delta\omega$, the energy
inside the cavity (for a laser resonant to the cavity) is
$W=P\eta/ \Delta\omega$. For an off-resonant cavity, where the
laser is detuned from the cavity by $\Delta$, the previous
expression for energy is multiplied by a Lorentzian:
\begin{equation}
f=\frac{1}{1+(2\Delta/\Delta\omega)^2}
\end{equation}
where $\Delta=\omega-\omega_0$ with $\omega_0$ being the resonance
frequency of the cavity. The energy in the cavity can also be
expressed as $\epsilon |E_{max}|^2 V_m$, where $\epsilon$ is the
permittivity of the medium, and $E_{max}$ is the electric field at
the point of maximum electric energy density, and $V_m$ is the
cavity mode volume. Equating the two expressions of energy, we can
write
\begin{equation}
\frac{P\eta} {\Delta\omega}
\frac{1}{1+(2\Delta/\Delta\omega)^2}=\epsilon |E_{max}|^2 V_m
\end{equation}
Using
\begin{equation}
\Delta\omega=\frac{\omega_0}{Q}=\frac{2\pi c}{Q\lambda_0}
\end{equation}
where $c$ is the velocity of light and $\lambda_0$ is the
resonance wavelength of the cavity, we can find that $E_{max}$:
\begin{equation}
\label{eqn_power_field} |E_{max}|=\sqrt{\frac{\eta P Q
\lambda_0}{2\pi c \epsilon
V_m}\frac{1}{1+(2\Delta/\Delta\omega)^2}}
\end{equation}
If the quantum dot is not located at the point of the maximum
electric field energy density, the electric field at its location
will be smaller than $E_{max}$ (and the spatial variation of the
E-field is determined by the mode pattern $\psi(x,y)$). Therefore,
the electric field at the location of the QD would be
\begin{equation}
|E_{cav}|=|E_{max}|\psi(x,y)
\end{equation}

On the other hand, when there is no cavity present, the intensity
$I$ of the light (assuming a Gaussian beam) incident on the GaAs
is given by
\begin{equation}
I=\frac{P}{2\pi \sigma_0^2}
\end{equation}
where $\sigma_0$ is the Gaussian beam radius of the laser. Also
the intensity of the laser is given by
\begin{equation}
I=\frac{1}{2}c\epsilon |E|^2
\end{equation}
Equating these two, the electric field is found to be
\begin{equation}
|E|=\sqrt{\frac{P}{c\epsilon\pi \sigma_0^2}}
\end{equation}
Assuming normal incidence on the air-GaAs interface
\begin{equation}
|E_{GaAs}|=\frac{2}{1+n}|E|
\end{equation}
where $n$ is the refractive index of GaAs. We note that the effect
of the reflection in the interface, is embedded in $\eta$ for the
analysis done for the cavity. From the above discussion, the
electric field sensed by the QD in the absence of the cavity has
the form
\begin{equation}
|E_{nocav}|=\frac{2}{1+n}\sqrt{\frac{P}{c\epsilon\pi \sigma_0^2}}
\end{equation}

Comparing the cavity and no-cavity case, we can find that the
electric field enhancement is given by
\begin{equation}
\frac{E_{cav}}{E_{nocav}}=\frac{1+n}{2}\sqrt{\frac{\eta Q
\lambda_0
W_0^2}{2V_m}\frac{1}{1+(2\Delta/\Delta\omega)^2}}\psi(x,y)
\end{equation}
When the laser is resonant with the cavity, the maximum field
enhancement for a linear three hole defect ($L_3$) cavity is $\sim
350$, assuming $\eta=1\%$, $Q=10000$,$\lambda_0=927$ nm,
$\sigma_0=3 \mu$ m, $V_m=0.8(\lambda_0/n)^3$, and the QD at the
field maximum, i.e., $\psi=1$. For a detuning of $4$ linewidths
(as is true for our experiment), the maximum enhancement is $\sim
40$. We note that this maximum enhancement can be increased by
using a better quality factor cavity, or lower mode volume.
Another way to increase the enhancement is increasing the coupling
efficiency $\eta$ by using a waveguide or a fiber coupled to the
cavity.
\section{Estimation of the QD Dipole Moment}
The data of Fig. $3$ a, b allows for order of magnitude estimation
of system parameters such as QD dipole moment and effective QD
electric field. Assuming a coupling efficiency of the Gaussian
laser beam to the PC cavity mode $\eta$, we can estimate the
maximum laser field amplitude $E$ at the position of the QD using
the Eqn. \ref{eqn_power_field}. From the linear fit in Fig. $3$ b,
we estimate the dipole moment $\mu_d$ of the QD to be be on the
order of $22$ Debye, with $\eta=1\%$ as obtained previously with
the same grating coupled cavity design
\cite{englund_toishi_perturbed_cavity}. For this dipole moment,
the maximum QD-cavity interaction strength $g/2\pi$ should be
$\sim 29$ GHz, assuming the QD is located at the electric field
maximum, thereby leading to the strong coupling. As mentioned
previously, we did not observe the anti-crossing of the QD and
cavity peaks in PL and thus believe that the actual value of $g$
is smaller than this calculated value most likely because the QD
is not located at the cavity electric field maximum.

\section{Theory of Bichromatic Driving}
In the theoretical description of our experiment, we calculate the
emission spectrum of the cavity under bichromatic driving of an
off resonantly coupled QD and measure how the intensity of the
cavity emission changes as a function of the probe laser detuning.
The bichromatic driving of a two-level system has been analyzed
before \cite{bichromatic_QD}. We use similar techniques to analyze
the driving of a two-level system such as a QD, coupled to an
off-resonant cavity. The dynamics of a driven QD-cavity system is
given by the Jaynes-Cummings Hamiltonian:
\begin{equation}
H=\omega_{cav} a^\dag a + \omega_{QD}\sigma^\dag \sigma +
g(\sigma^\dag a + \sigma a^\dag) + J\sigma + J^*\sigma^\dag
\end{equation}
where $\omega_{cav}$ and $\omega_{QD}$ are, respectively, the
cavity and the dot resonance frequency; $a$ and $\sigma$ are,
respectively, the annihilation operator for a cavity photon and
the lowering operator for the QD; $g$ is the coherent interaction
strength between the QD and the cavity and $J$ is the Rabi
frequency of the driving laser. For bichromatic driving, the
driving field $J$ consists of a strong pump laser with Rabi
frequency $J_1$ tuned to the QD frequency and a weak probe laser
with Rabi frequency $J_2$, which can be tuned to arbitrary
frequency, parameterized by the pump-probe detuning $\delta$:
\begin{equation}
J=J_1 e^{i\omega_{QD} t} + J_2 e^{i (\omega_{QD}+\delta) t}
\end{equation}

In a frame rotating with the pump laser frequency the Hamiltonian
is
\begin{equation}
  H = H_0+H(t)
   =  \Delta a^\dag a + g(\sigma^\dag a + \sigma a^\dag) + J_1\sigma_x
+ J_2\left(e^{i\delta t}\sigma + e^{-i\delta
t}\sigma^\dagger\right)
\end{equation}
where $\Delta=\omega_{cav}-\omega_{QD}$ is the QD-cavity detuning.
We note that for a bichromatic driving, the Hamiltonian is always
time-dependent. To treat incoherent processes we use the master
equation \cite{book:quan_noise}:
\begin{eqnarray*}
  \dot{\rho} &=& -i [H_0+H(t),\rho] +
\mathcal{D}\left(\sqrt{2\gamma}\sigma\right)\rho +
\mathcal{D}\left(\sqrt{2\kappa}a\right)\rho +\\
   & & \mathcal{D}\left(\sqrt{2\gamma_r\bar{n}}a^\dagger\sigma\right)\rho
+\mathcal{D}\left(\sqrt{2\gamma_r(1+\bar{n})}a\sigma^\dagger\right)\rho
+ \mathcal{D}\left(\sqrt{2\gamma_d}\sigma^\dagger\sigma\right)
\end{eqnarray*}
where $\mathcal{D}\left(C\right)\rho$ is the Lindblad term $C\rho
C^\dagger-\frac{1}{2}\left(C^\dagger C\rho + \rho C^\dagger
C\right)$  associated with the collapse operator $C$. The first
two terms represent QD spontaneous emission with a rate $2\gamma$,
and cavity decay with a rate $2\kappa$. The two terms with
$\gamma_r$ represent a phonon mediated coupling between the cavity
and the QD \cite{majumdar_phonon_11}. The last term with
$\gamma_d$ phenomenologically describes pure dephasing of the QD.
We numerically calculate the emission spectrum of the cavity given
by the Fourier transform of the two-time correlation function of
the cavity field, proportional to $\langle a^\dagger (\tau)
a(0)\rangle$. Under the quantum regression theorem the
auto-correlation function is equal to tr$\{ a^\dagger M(\tau)\}$
where $M(\tau)$ obeys the master equation with initial condition
$a\rho(t\rightarrow\infty)$. The time dependence of the
Hamiltonian is such that the master equation can be cast in terms
of Liouvillian superoperators as
\begin{equation}\label{rhoEOM}
\dot{\rho}=\left(\mathcal{L}_0+\mathcal{L}_+e^{i\delta
t}+\mathcal{L}_-e^{-i \delta t}\right)\rho
\end{equation}
where
\begin{eqnarray*}
  \mathcal{L}_0\rho &=&  -i [H_0,\rho] +
\mathcal{D}\left(\sqrt{2\gamma}\sigma\right)\rho +
\mathcal{D}\left(\sqrt{2\kappa}a\right)\rho\\
&&+\mathcal{D}\left(\sqrt{2\gamma_r\bar{n}}a^\dagger\sigma\right)\rho
+\mathcal{D}\left(\sqrt{2\gamma_r(1+\bar{n})}a\sigma^\dagger\right)\rho
+ \mathcal{D}\left(\sqrt{2\gamma_d}\sigma^\dagger\sigma\right)\\
  \mathcal{L}_+\rho &=& -i[\sigma,\rho] \\
  \mathcal{L}_-\rho &=&-i[\sigma^\dag,\rho]
\end{eqnarray*}
This equation is solved by Floquet theory, by assuming a solution
of the form
$\rho(t)=\displaystyle\sum_{n=-\infty}^{\infty}{\rho_n(t) e^{i
n\delta t}}$. The number of terms in the expansion necessary to
obtain any level of precision is determined by the relative
strength of $J_1$ to $J_2$, and in this way the problem can be
considered perturbative in the probe strength. After Laplace
transforming the master equation, the method of continued
fractions is used to obtain the resonance fluorescence spectrum of
the cavity \cite{qotoolbox}. The height of the peak at the cavity
resonance is calculated as a function of the probe detuning
$\delta$. The criterion for the appearance of dressed states is
that the pump Rabi frequency $J_1$ should be higher than the QD
linewidth $2\gamma$. The inclusion of incoherent terms $\gamma_r$
and $\gamma_d$ effectively broadens the dot and alters this
condition, but below a certain critical value of $J_1$ the change
in the cavity height with probe detuning is a simple Lorentzian
with a linewidth on the order of the natural QD linewidth. Above
threshold, the dressed states are resolvable and cavity height
spectrum splits into two peaks in the experimental regime we
considered. Broadening of the peaks in the experiment beyond
theoretical prediction is caused by the spectral diffusion of the
QD. The parameters used for the simulations are: $\kappa/2\pi=17$
GHz, $\gamma/2\pi=1$ GHz, $\gamma_r/2\pi=.5$ GHz,
$\gamma_d/2\pi=3$ GHz, $\Delta=8\kappa$, $\bar{n}=1$. For the
simulation reported here we assume $g=0$, as the QD is not
strongly coupled to the cavity. Increasing $g$ makes the two peaks
more asymmetric.

\section{Numerical Simulation: Dependence of the splitting on pump power}
We show in the paper that we can probe the coherent interaction
between the QD and the resonant laser by monitoring the
off-resonant cavity emission (Fig. $1$ c in the paper). Both
theoretically and experimentally we observe two peaks at lower
pump powers. At higher pump power, we theoretically observe two
dips. In our experiment, however, we cannot reach this regime of
high pump power.

\begin{figure}
\centering
\includegraphics[width=3in]{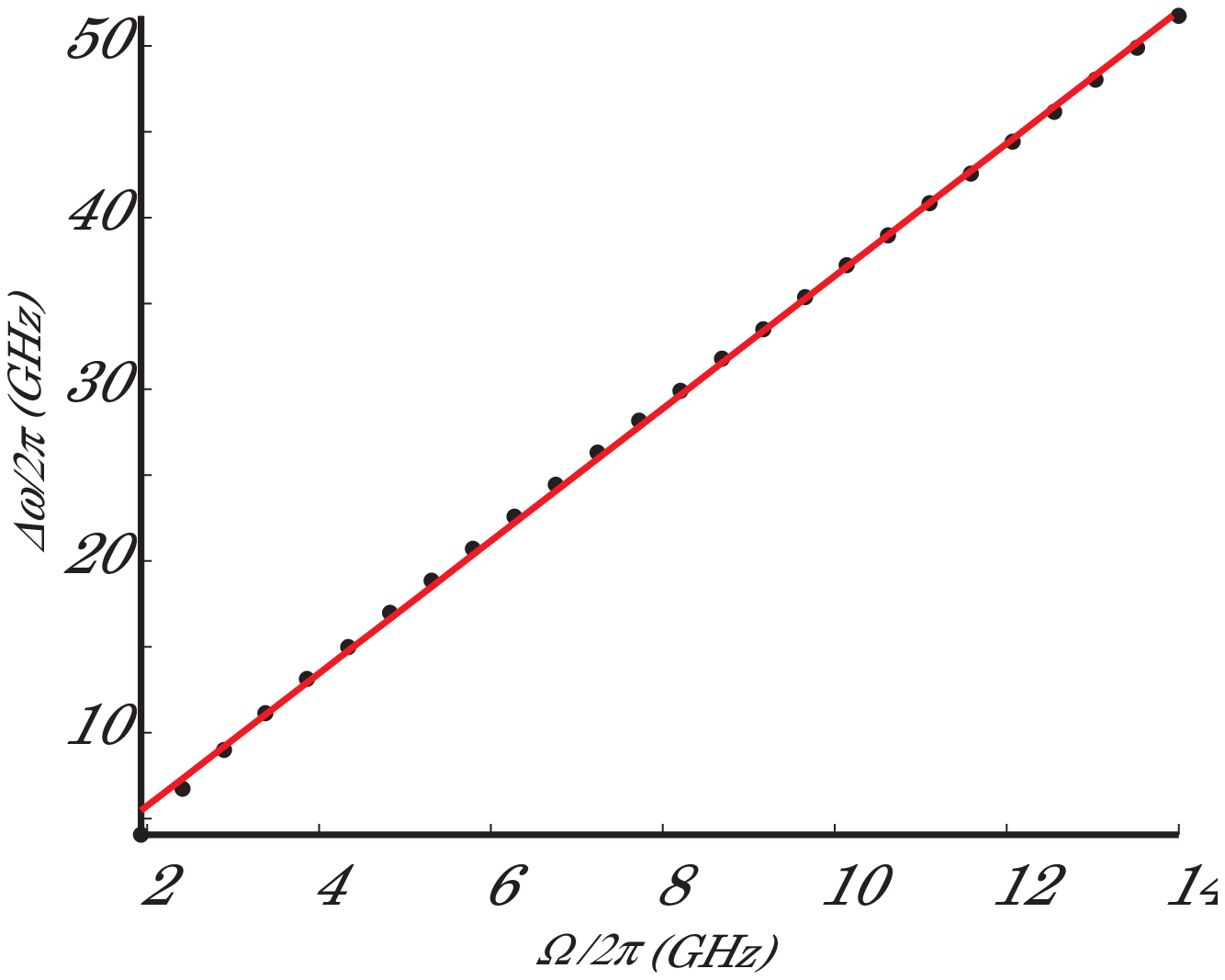}
\caption{ The separation between the two peaks (as shown in Fig.
$1$ c in the paper) as a function of the laser Rabi frequency. The
slope of the linear fit is $\sim 4$.\label{Fig_suppl_peaks}}
\end{figure}

\begin{figure}
\centering
\includegraphics[width=3in]{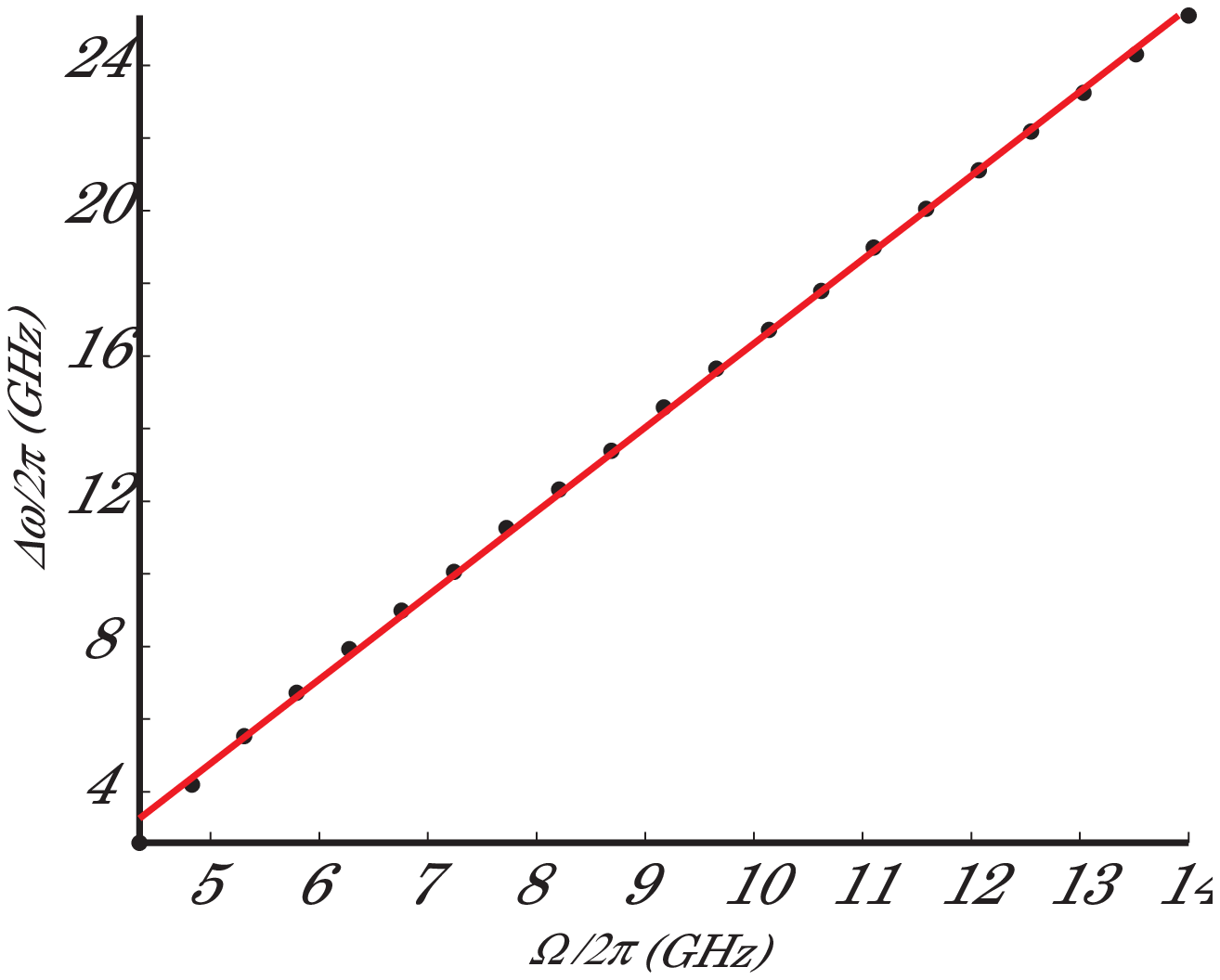}
\caption{ The separation between the two dips (as shown in Fig.
$1$ c in the paper) as a function of the laser Rabi frequency. The
slope of the linear fit is $\sim 2$. \label{Fig_suppl_dips}}
\end{figure}

The separation between the peaks and dips increases linearly with
the pump Rabi frequency. However, from the theoretical plot, we
find that the peaks are separated by $4$ times the laser Rabi
frequency (Fig. \ref{Fig_suppl_peaks})and the dips are separated
by twice the laser Rabi frequency (Fig. \ref{Fig_suppl_dips}).
More detailed theoretical derivation will be provided in
\cite{papag_prep}.

\section{Numerical Simulation: effect of $g$}
\begin{figure}
\centering
\includegraphics[width=3in]{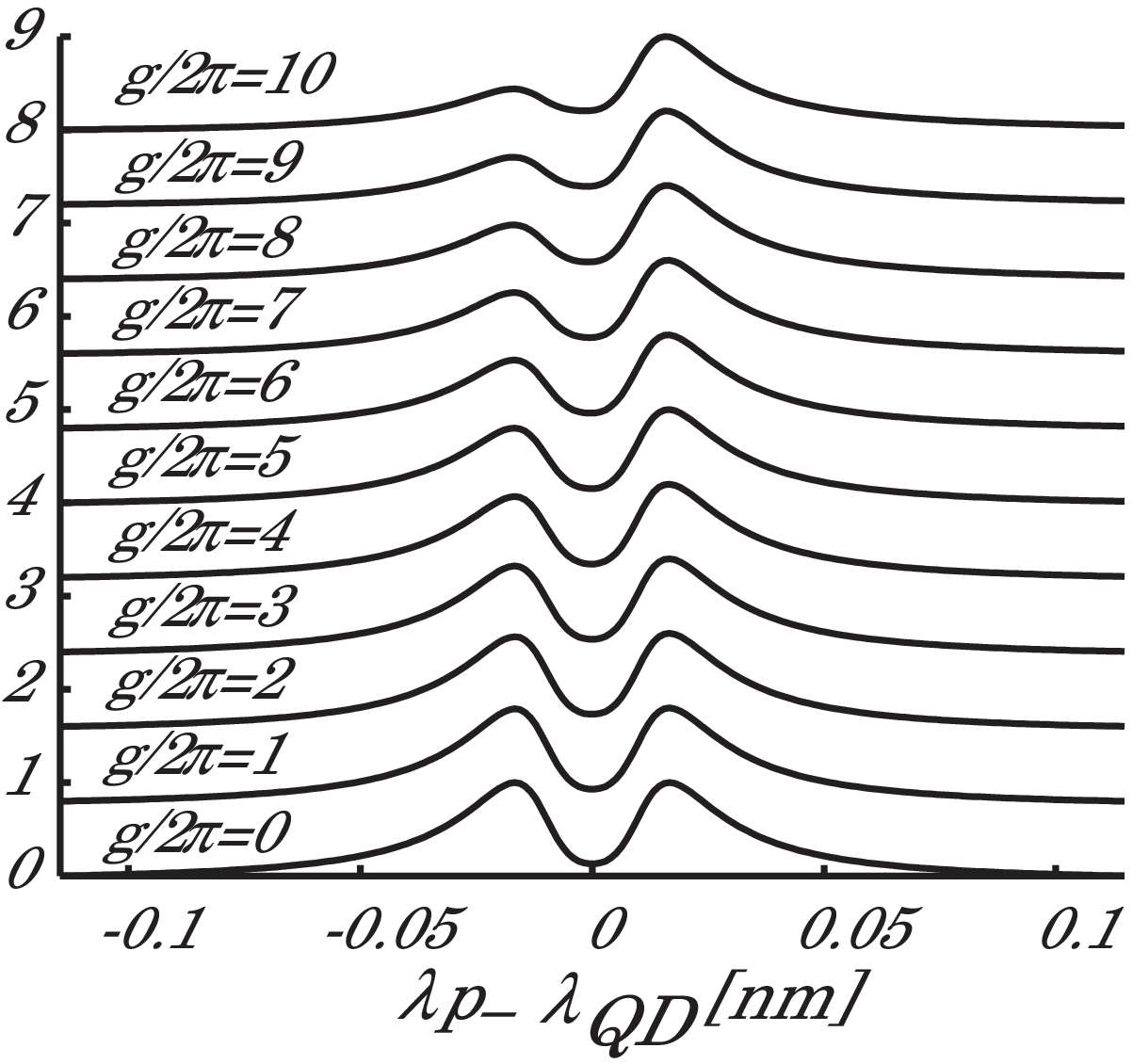}
\caption{ Cavity emission as a function of probe laser wavelength,
for different dot-cavity coupling $g$.
\label{Fig_suppl_effect_of_g}}
\end{figure}

In the numerical simulation results presented in the paper, we
assumed $g=0$, i.e., no coherent interaction is present between
the QD and the cavity. Inclusion of $g$ makes the two peaks
asymmetric. Fig. \ref{Fig_suppl_effect_of_g} shows the cavity
emission for a pump power of $25$, with $g/2\pi$ ranging from $0$
to $10$. Here the cavity is at a shorter wavelength compared to
the QD, and we observe that the peak closer to cavity is not
enhanced. This observation is starkly different from the resonance
fluorescence measurement, where the peak close to the cavity is
enhanced, as observed in \cite{majumdar_phonon_11,hughes_mollow}.
This indicates again, that this way of measuring the coherent
interaction between the QD and the laser is akin to an absorption
measurement.

The authors acknowledge financial support provided by the Army
Research Office, Office of Naval Research and National Science
Foundation. A.M. was supported by the Stanford Graduate Fellowship
(Texas Instruments fellowship). E.K. was supported by the
Intelligence Community (IC) Postdoctoral Research Fellowship.
\bibliography{NRDC_bibl_PR}
\end{document}